\scrollmode
\documentstyle[draft]{mn}

\input epsf

\title[Helium white dwarf]{Helium white dwarf in PSR J0751+1807; too cool,
in PSR J1012+5307; too hot?}

\author[Ene Ergma et al.]{ Ene
Ergma$\rm ^1$, Marek J. Sarna$\rm ^2$ and Jelena 
Ger\v{s}kevit\v{s}--Antipova$\rm^1$ \\
$\rm ^1~$Physics Department, Tartu University, \"Ulikooli 18, EE2400
       Tartu, Estonia \\
$\rm  ^2~$N. Copernicus Astronomical Center,
       Polish Academy of Sciences,
       ul. Bartycka 18, 00--716 Warsaw, Poland. \\}

\date{\small Accepted . Received ; in original
form 1999}

\begin{document}

\maketitle

\begin{abstract}

We discuss the cooling history of the  low--mass, helium core white dwarfs
in  short orbital period   millisecond pulsars
PSR J0751+1807 and PSR J1012+5307. The revised cooling age estimated by
Alberts et al. agrees with the age estimation for PSR J1012+5307, removing
the discrepancy between the spin--down age and the cooling age. However,
if we accept this model then the helium white dwarf in the
binary pulsar system PSR J0751+1807 must be much hotter than is observed.
We propose that this discrepancy may be resolved if, after detachment
of the secondary star from its Roche lobe in PSR J0751+1807, the star loses
its hydrogen envelope due to pulsar irradiation. When hydrogen burning
stops, the white dwarf will cool down much more quickly than in the case
of a thick hydrogen envelope with a hydrogen burning shell. We discuss
several possibilities to explain different cooling histories of  
 white dwarfs in both  systems. 

\end{abstract}


\section{Introduction}

It is accepted that formation of a binary millisecond (or recycled)
pulsar with a low--mass companion may be
explained as the end--point of close binary evolution in which an old
pulsar is spun--up by accretion from the red giant (secondary)
(see Bhattacharaya \& van den Heuvel 1991, for review). Once
the secondary overflows its Roche lobe, it starts transferring mass and
angular momentum to the neutron star. During the long term accretion phase
the neutron star is generally assumed to be spun--up to the equilibrium 
period ($P_{eq}$).
The equilibrium period can be written as (Bhattacharaya \& van den Heuvel 
1991):

\begin{equation}
P_{eq}=1.9~ms~ B_9^{6/7}\left(\frac{M_{ns}}{1.4M_\odot}\right)^{-5/7}\left
(\frac{\dot{M}}{\dot{M}_{Edd}}\right)^{-3/7} R_6^{16/7}~,~
\end{equation}

where $B_9$ = B/$10^9$ G, $R_6$= $R_{ns}$/$10^6$ cm and $M_{ns} $ are
the magnetic field strength, radius and mass of the neutron star respectively and ${\dot M}_{Edd} $
is the Eddington mass accretion rate.
After the whole envelope of the red giant has been transferred, the
binary consists of a low--mass star with a helium core (progenitor of a
low--mass helium white dwarf) and a fast rotating neutron star (millisecond
pulsar). At this point the pulsar spin period starts to change due to
magneto--dipole radiation, with a characteristic time {\it t} elapsed from
this epoch
\begin{equation}
{\it t} = \tau \times [\frac{2}{n-1}][1- (\frac{P_0}{P})^{n-1}]~,~
\end{equation}
where $\tau$=P/(2$\dot{P}$), $P_0$ and P  are the initial and
current spin periods, and $\dot{P}$ is the spin period 
derivative. n is the braking index (=3) and if $P_0\ll$P then {\it t}=$\tau$.
It is possible to estimate the age {\it t}
from the cooling age of the white dwarf. The latter can be calculated
from the observed parameters of the ($T_{eff}$, $\log ~g$). This gives 
a unique opportunity for  determination of
neutron star age which is independently of the spin--down theory of pulsars.

In this short note we shall discuss  cooling properties of  helium
white dwarfs in  short period  binary systems with millisecond pulsars,
PSR J0751+1807 and PSR J1012+5307.

\section{Observational data for millisecond pulsars}

For both systems, the characteristic time {\it t} (eq (2)), to be compared 
with the cooling age of the white dwarf, is assumed to be equal to the 
spin--down age 
$\tau$ of the neutron star. This requires $P_0\sim P_{eq}\ll P$. The last 
restriction will be relaxed in the discussion section (par. 6).

\subsection{PSR J0751+1807}

This system consists of a 3.5 ms pulsar in a 6 h orbit with a $\sim$ 0.15
$M_\odot$ companion (Lundgren, Zepka \& Cordes 1995). We estimate the 
magnetic field strength of neutron star from the relation

\begin{equation}
B= 3\times 10^{19}(P\dot{P})^{1/2}~.
\end{equation}

For the observed $P$ and $\dot{P}$ ($8 \times 10^{-21} ~ss^{-1} $) we have
B $\sim$ 1.7$\times 10^8$ G, and the spin--down
age of the neutron star, $\tau$ $\sim ~$8 Gyr. For this age and 
$M_{wd} \sim 0.15$ $M_\odot$ a simple white dwarf cooling model
by Mestel (1952) gives  L
$\sim 10^{-4.3}L_\odot$. Lundgren at al. (1996) estimated a lower bound for
the visual magnitude, $m_{\rm v} >$23.5, which requires $T_{eff}<$
9000 K and an age greater than $7.94 \times 10^{8}$ years. From their data
a main-sequence companion is ruled out down to an M5 star
($M\sim$ 0.10 $M_\odot$), since
such a  star  would have been detected at 
the distance of the pulsar.  Therefore the pulsar 
companion in this system is clearly a white dwarf.

\subsection{PSR J1012+5307}

PSR J1012+5307 consists of a millisecond pulsar (with $P~$= 5.256 ms)
in a 14.52 h orbit
with a $\sim$ 0.15 $M_\odot$ companion ($M_{ns}$=1.4 $M_\odot$,
${\it i}=60^0$).
$\dot{P}$=1.46$\times 10^{-20}ss^{-1}$  leads to the spin--down age, $\tau
\sim$ 7.0 Gyr (Nicastro et al. 1995, Lorimer et al. 1995), and
B $\sim$ 2.6 $\times 10^8$ G. Van Kerwijk et al. (1996) and
Callanan et al. (1998) estimated the mass of the helium white dwarf to be 0.16
$\pm$ 0.02 $M_\odot$.
Taking the
observationaly determined gravity and effective temperature for the white
dwarf, and
adopting the cooling model of Iben \& Tutukov (1986), the cooling age has
been estimated as 0.3 Gyr (Lorimer et al. 1995). The discrepancy between
these two age estimations is solved if, in the white dwarf cooling models,
a thick hydrogen envelope with stationary hydrogen shell burning is taken
into account (Alberts et al. 1996, Driebe et al. 1998), rising the cooling
age to several Gyr.

\section{The evolutionary code}

The evolutionary sequences we have calculated consist of
three main phases:

\noindent
(i) detached evolution lasting until the companion
fills its Roche lobe,

\noindent
(ii) semi--detached evolution,

\noindent
(iii) a cooling phase of the white dwarf (the final phase during which a
system with a ms pulsar + low--mass helium white dwarf is left behind).

\subsection{Detached phase}

The duration
of the detached phase is somewhat uncertain;
it may be determined either by the nuclear time--scale or by the
much shorter time--scale of the orbital angular momentum loss owing
to the magnetized stellar wind.

\subsection{Semi--detached phase}

In our calculations we assume that the semi--detached evolution of a binary
system is non--conservative, i.e. the total mass and angular momentum of the
system are not conserved. The formalism which we have adopted is described in
Muslimov \& Sarna (1993). We introduce the
parameter, $f_1$, characterizing the loss of mass from the binary system
and defined by the relations,

\begin{equation}
\dot M = \dot M_2 f_1 ~~~~and~~~~ \dot M_1
= - \dot M_2 (1 - f_1) ,
\end{equation}

where $\dot M$ is the mass--loss rate from the system, $\dot M_2$ is
the rate of mass--loss from the  secondary star, and $\dot M_1$ is
the accretion rate onto the neutron star (primary). The matter leaving the
system will remove  angular momentum
according to the formula

\begin{equation}
\left. {{\dot J} \over {J}} \right|_{ML} = f_2 {{M_1 \dot M} \over {M_2 M}}
~~~~~yr^{-1}  ,
\end{equation}

where $M = M_1$+$M_2$, $M_1$ and $M_2$ are the masses of the neutron star 
and secondary star respectively. Here we have introduced an
additional parameter,
$f_2$, which describes the efficiency of the orbital angular momentum
loss from the system due to a stellar wind (Tout \& Hall 1991). In our
calculations we assume $f_1$=1 and $f_2$=1.

We also assume that the secondary star, possessing a convective envelope,
experiences magnetic braking (Mestel 1968; Mestel \& Spruit 1987; Muslimov \&
Sarna 1995), and as a consequence of this, the system loses its orbital
angular momentum. For magnetic stellar wind we used the formula for the
orbital angular momentum loss
\begin{equation}
\left. {{\dot J} \over {J}} \right|_{MSW}= -3\times 10^{-7}{{M^2R_2^2} \over 
{M_1M_2 a^5}}~~~~~yr^{-1}  ,
\end{equation}

where $a$ and $R_2$ are the separation of the components and the radius of 
the secondary star in solar units.

In the semi--detached phase we have also included the effect of illumination
of the secondary
star by the millisecond pulsar. In our calculations we assume that the
atmosphere of the secondary star is heated  by the hard (X--ray and $\gamma$--ray)
radiation from the pulsar 
 (Muslimov \& Sarna 1993). The effective temperature,
$T_{eff} $, of the companion during the illumination stage is
determined from the relation

\begin{equation}
L_{in} + P_{ill} = 4 \pi \sigma R_2^2 T^4_{eff} ,
\end{equation}

where $L_{in} $ is the intrinsic luminosity corresponding to the
radiation flux coming from the stellar interior, $\sigma $ is the 
Stefan--Boltzmann constant and $ R_2 $ is the  radius of the secondary. 
$P_{ill} $ is the millisecond pulsar radiation that heats the photosphere,
given by
\begin{equation}
P_{ill}= f_3 \left(\frac{R_2}{2a}\right)^2 L_{rot},
\end{equation}

where $L_{rot}$ is ``rotational luminosity'' of the neutron star due to 
magneto--dipole radiation (plus a wind of relativistic particles)

\begin{equation}
L_{rot}=\frac{2}{3c^3} B^2 R_{ns}^6\left(\frac{2\pi}{P}\right)^4
\end{equation}

in the above formula $R_{ns}$ is the neutron star radius, 
$f_3$ is a factor introduced to account for
the efficiency of transformation of the irradiation flux into thermal energy 
(in our case we take $f_3 = 2\times 10^{-3}$ Muslimov \& Sarna (1993)). 
Note that in our calculations
the effect of irradiation is formally treated by means of modification
of the outer boundary condition, according to the above relation.

\subsection{Cooling phase}

For systems in the white dwarf cooling phase, 
 we take into account two processes:

\noindent
(i) the loss of
orbital angular momentum due to the emission of gravitational radiation
(Landau \& Lifshitz 1971),

\noindent
(ii)  an irradiation induced stellar wind, with corresponding loss of
orbital angular momentum estimated as

\begin{equation}
\left. {{\dot{J}} \over {J}} \right|_{wind}=\frac{q}{1+q}\frac{\dot{M}_{wind}}{M_2}~,
\end{equation}

where $q$ is the mass ratio ($q=M_1$/$M_2$) and ${\dot M}_{wind} $ is a 
mass loss of an irradiation induced stellar wind defined by the simple van den
Heuvel \& van Paradijs (1988) model

\begin{equation}
\dot{M}_{wind}=f \frac{R_2}{GM_2}\left(\frac{R_2}{2 a}\right)^2\frac{2R_{ns}^6}
{3c^3} B^2\left(\frac{2\pi}{P}\right)^4~,
\end{equation}

where $f$ is efficiency factor (from 0 to 1) and $a$ is the separation
between the stars
in solar units. In our calculations we assume three values of $f$: 0, 0.1 and
0.5 (very efficient irradiation).

\subsection{The code}

The models of the stars filling their Roche lobes were computed using a
standard stellar evolution code based on the Henyey--type code of Paczy\'nski
(1970), which has been adapted to low--mass stars. Convection is treated with
the mixing--length algorithm proposed by Paczy\'nski (1969). We solve the
problem of radiative transport by employing the opacity tables of Iglesias \&
Rogers (1996). Where the Iglesias \& Rogers (1996) tables
are incomplete, we switch to opacity tables of Heubner et
al. (1977). For temperatures less than 6000 K we use  opacities calculated by
Alexander \& Ferguson (1994) and Alexander (private communication).
 Finally, we assume a Population I chemical composition for
the secondary (X=0.7, Z=0.02).

\section{Evolutionary considerations}

To produce short orbital period systems composed of a low--mass helium white
dwarf and a millisecond pulsar, it is necessary that the low--mass secondary
(red giant) fills its Roche lobe
below the so called bifurcation period, and above the boundary period
(Ergma \& Sarna 1996, Ergma, Sarna \& Antipova 1998). Ergma et al. (1998)
and recent calculations by Sarna, Ergma \& Antipova (2000) have shown
that after detachment of the secondary star from its Roche lobe, a rather thick
hydrogen layer is left
on the top of the helium core. The same result has been obtained by Driebe et
al. (1998). This thick hydrogen layer will greatly influence the cooling
history of the very low--mass helium white dwarf ($< 0.25 ~M_\odot $).
Alberts et al. (1996) were
the first to argue that because white dwarfs with $M_{wd}$$<$0.20 ~$M_\odot$ do
not show
thermal flashes (as found by Webbink, 1975),  the amount of hydrogen in
the envelopes must remain high, resulting in a much longer phase of significant
hydrogen burning. Thus   cooling times for these objects are considerably
extended.
This removes the discrepancy between the cooling  and spin--down ages for the
millisecond binary systems PSR J1012+5307 (Alberts et al. 1996, 
Driebe et al. 1998, Sarna et al. 1999,2000).

But now the question arises for the PSR J0751+1807. The evolutionary history
for both pulsar systems are very similar, as argued by Ergma \& Sarna (1996).
This means that in
PSR J0751+1807 the helium white dwarf must also be in the stable, long--lasting
 hydrogen--burning
phase, which leads to  long cooling times.
However, observations show that the white dwarf in this system is
undetectable, and really very old.

To solve this discrepancy, we propose that irradiation of the secondary star
by the millisecond pulsar PSR J0751+1807 causes  mass loss
after detachment from the Roche lobe. After detachment the star with a
hydrogen burning shell
evolves from the red giant region towards higher effective temperatures. 
If we accept the simple irradiation driven mass loss model proposed by van 
den
Heuvel \& van Paradijs (1988), then using eq. 11, for $R_2$= 0.56$~R_\odot$
(see point 2 on Fig. 1), $a = 2~R_\odot$, $M_2$=0.15$~M_\odot$,
$R_{ns}=10 ~km$ and $f=0.1$ (efficiency factor), we have
$\dot{M}_{wind} \sim $
4$\times 10^{-10} ~M_\odot yr^{-1}$. Since the evolutionary time--scale
to evolve  is rather long (at least several
hundred million years), about 0.01$~M_\odot$ of the hydrogen envelope will
be lost.
As a result, the hydrogen burning
shell will switch off, and the white dwarf may cool down very quickly.

\begin{figure}
\epsfverbosetrue
\begin{center}
\leavevmode
\epsfxsize=7cm\epsfbox{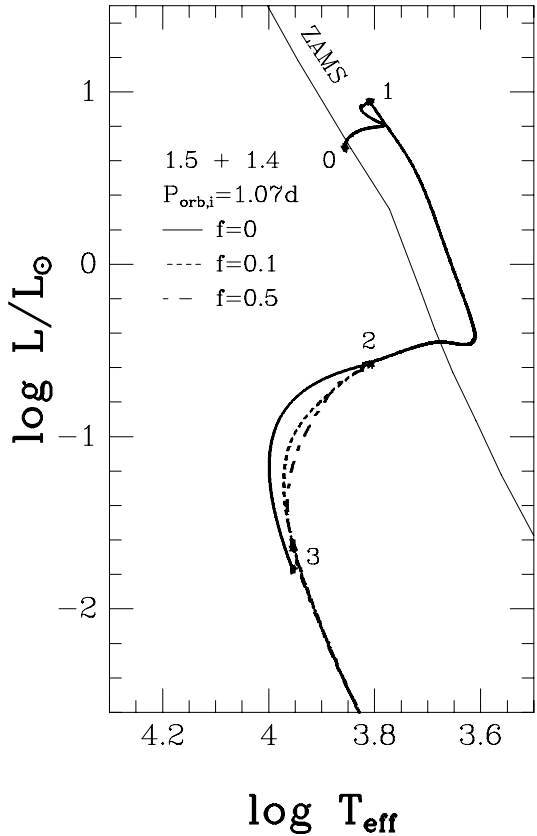}
\end{center}
\caption{HR diagram with evolutionary tracks of a 1.5$~M_\odot$ star with
1.4$~M_\odot $ neutron star. Different phases of
evolution are: detached evolution before filling the Roche lobe (0--1),
semi--detached evolution (1--2), and detached evolution during the cooling
of a white dwarf (2--3). Phase (2--3) is plotted for three different cases;
normal cooling
(thick solid line), cooling with irradiation induced wind (f=0.1 -- thick
dashed line) and very efficient irradiation induced wind (f=0.5 -- thick
dash--dotted line).
The thin solid line shows the ZAMS as calculated
using our code.}
\end{figure}

\section{Simple cooling model for the white dwarf in PSR J0751+1807}

To illustrate how the  mass loss due to irradiation will influence the cooling
evolution of the pre--white dwarf (after detachment from  Roche lobe), we
computed the evolutionary sequence for a system of 1.5 + 1.4 $M_\odot$, with
$P_{orb,i}$ (RLOF)=1.07 d (RLOF -- Roche lobe overflow) and Z=0.02 (Fig. 1).
After detachment from the Roche lobe,  mass loss at a rate estimated
from eq. 11 has been included (Fig. 2).

\begin{figure}
\epsfverbosetrue
\begin{center}
\leavevmode
\epsfxsize=7cm
\epsfbox{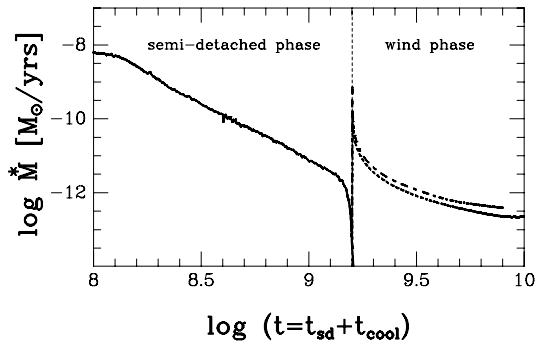}
\end{center}
\caption{The dependence of the mass loss rate as a function of time
in two different phases of evolution: semi--detached evolution (phase (1--2)
-- solid line) and cooling with irradiation induced wind. For the wind phase
we show two different efficiencies of the wind: f=0.1 (dashed line) and f=0.5
(dash--dotted line).}
\end{figure}

The  mass loss phase leads to a change of the orbital angular momentum, which
leads to significant changes in the orbital period evolution (Fig.3). Figure 4
shows the evolution of the visual magnitude ($m_V$) of the progenitor of the
white dwarf during the cooling phase, with wind mass loss included.

\begin{figure}
\epsfverbosetrue
\begin{center}
\leavevmode
\epsfxsize=7cm
\epsfbox{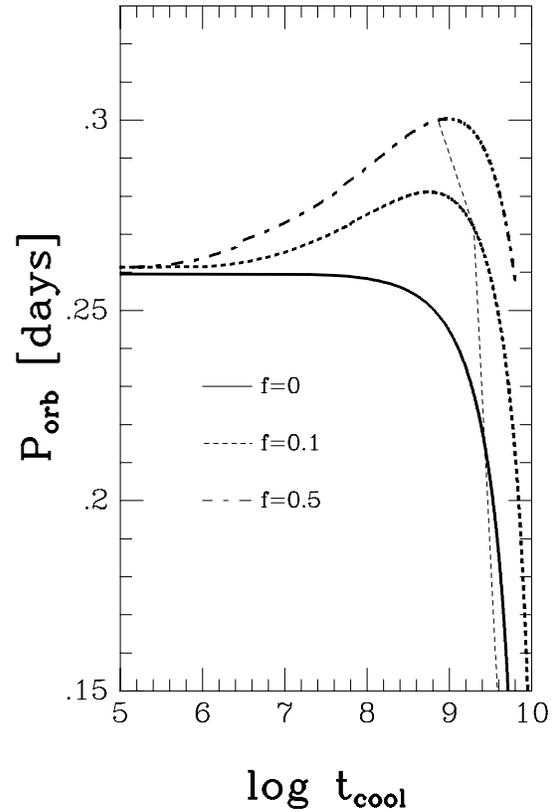}
\end{center}
\caption{Evolution of the orbital period during the cooling phase for the
three cases considered. The meaning of the lines is the same as for Fig. 1.
The thin dashed line shows location, at which the surface effective temperature of the
white dwarf is equal to 9000 K.}
\end{figure}

\begin{figure}
\epsfverbosetrue
\begin{center}
\leavevmode
\epsfxsize=7cm
\epsfbox{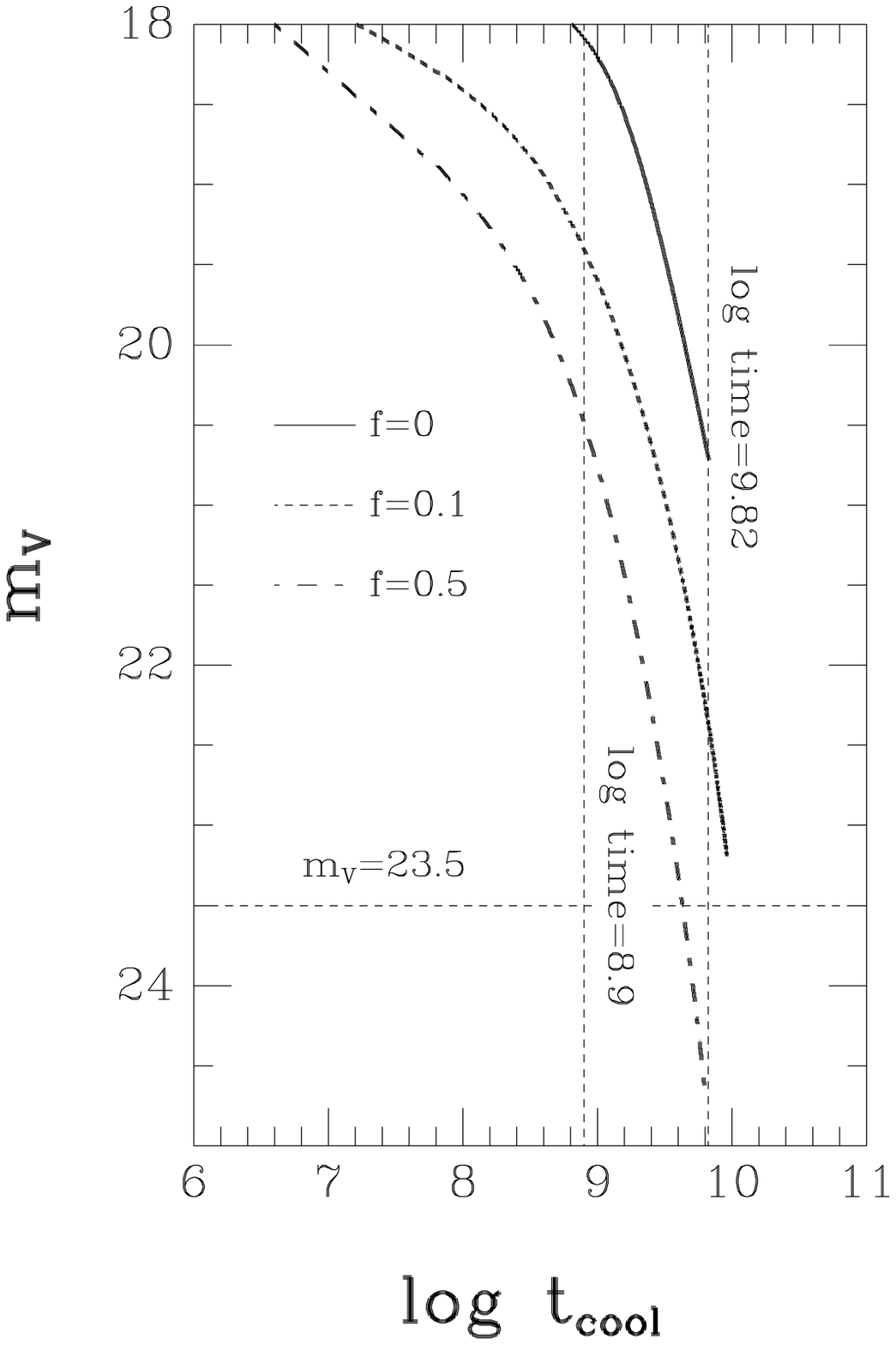}
\end{center}
\caption{Evolution of the visual magnitude during the cooling phase for the
three cases considered. The meaning of lines is the same as for Fig. 1.
The thin horizontal dashed line shows the Lundgren et al. (1996) lower bound
for visual magnitude. The two thin vertical dashed lines  show lower and
upper limits for the millisecond pulsar age.}
\end{figure}

In the H--R diagram (Fig. 1) the evolution
of this system is presented with and without 
 mass loss. The decrease of the mass of the hydrogen envelope
leads to a significant decrease of the hydrogen shell burning efficiency.
Therefore, if compared to the evolution of a white dwarf with stationary
hydrogen shell burning, the cooling time is decreased.
For example, when log L/$L_\odot$ = --2.0, the cooling age without mass loss
is 8.4$~ \times 10^9 ~$yrs, with moderate mass  loss
(f=0.1) is 5.07$~ \times 10^9 ~$yrs, and with  very efficient mass loss (f=0.5)
is 2.17$~ \times 10^9 ~$yrs.
Due to angular momentum loss by the irradiation induced wind, the orbital
period and separation of the system (after detachment of the secondary
from the Roche lobe) initially increases (Fig. 3). Because of this increase
in the separation, and shrinking of the white dwarf  (see eq. 11)
the  mass loss rate (Fig. 2) decreases by two orders of magnitude,
making this mechanism of orbital angular momentum loss insignificant.
The orbital period starts to decay due to emission of gravitational waves. 

Figure 4 gives the limit for acceptable $m_V$ from theoretical cooling
tracks. We can exclude normal cooling (with shell hydrogen burning)
due to a too low $m_V$. Two other cases of cooling with irradiation induced
wind give the right range of visual magnitude. For the upper limit of 
pulsar age (spin--down age $\sim ~$8 Gyrs)  $m_V$ and $T_{eff}$ are
(22.4, 7200K) for $f=0.1$ and (24.7, 4700K) for $f=0.5$.

One can ask why we do not observe circumbinary matter inside PSR
J0751+1807, as we do in PSR 1957+20 or PSR 1744--24A (Lyne 1995).
We propose that after having lost most of its hydrogen envelope
(during the high mass loss phase),the star hydrogen burning stops   
and relaxes to a low--mass helium white dwarf. Irradiation of the white dwarf
by the pulsar may be significant, but it does not lead to change in the
internal structure of the envelope (as it does for a low--mass star with
a convective envelope and shell nuclear burning). Therefore, no hot extended
envelope is created and no mass loss induced.

\section{Discussion}

The first question is what is the real value of the initial pulsar period 
$P_0$ for both systems? Is it equal to $P_{eq}$? To explain the discrepancy 
between the cooling and spin--down ages of PSR J1012+5307 
Burderi et al. (1996) have accepted the 
so called accretion--induced magnetic field decay model 
(see Bhattacharya \& Srinivasan 1995 for review). In this case, at the end
of the 
accretion phase, field decay takes place on a shorter time--scale than 
spin--up, and spin equilibrium is no longer achieved.

The second point is connected with the  discussion of Webbink's (1975) 
results, who found that there is a hydrogen burning in the thick envelope 
(of a few $\times10^{-3} $$M_\odot$) of low--mass helium white dwarfs 
($< 0.2 ~M_\odot$). Recently three different groups, Albert et al. (1996, 
using Eggleton evolutionary program), Driebe et al. 1998 (Kippenhahn program)
and we (Sarna et al. 1999,2000, Paczy\'nski program) have obtained the same 
results: small mass helium white dwarfs ($<$ 0.2$M_\odot$), after detachment 
from the Roche
lobe, have a massive hydrogen layer (a few $\times 10^{-2}~M_\odot$) in which
stationary hydrogen burning continues. This keeps these low--mass helium
white dwarfs hot and prolongs their cooling time.

However, there is a difference between Alberts et al. and our+Driebe's results, 
since we and Driebe found hydrogen flashes for M$>$ 0.2 $M_\odot$ and 
Alberts et al. did not.
We show that the transition from a cooling white dwarf with a temporarily 
stable hydrogen--burning shell to a cooling white dwarf in which almost all 
residual hydrogen is lost in a few thermal flashes (via Roche--lobe overflow)
occurs between 0.183--0.213 $M_\odot$, which well agrees with the cooling 
scenario proposed by Burderi et al. (1998).

Therefore we may accept that PSR J1012+5307 is in a long term hydrogen shell 
burning stage and this may be the reason why the white dwarf is too hot 
(accepting that spin--down age of 7 Gyr, which value could be even larger if 
the pulsar has a significant transverse velocity (Hansen \& Phinney 1998), 
is almost equal to the cooling age). For PSR J0751+1807
the irradiation driven mass loss reduces the hydrogen envelope mass,
the hydrogen burning deceases and the white dwarf is quickly cooled.

Taking into account the effect of illumination, $P_{eq}$ is 
$\sim 1.6\times 10^{-3}$s (Ergma \& Sarna 1996). Therefore, if the pulsar 
in PSR J1012+5307 has been spun up to $P_{eq}$, it also has suffered 
irradiation driven mass loss. If we do not include 
illumination then $P_{eq}$ is longer, since at the beginning of 
RLOF mass transfer is proceeding on the thermal time--scale at a rather high mass 
accretion rate, and the neutron star spins--up to millisecond periods. After 
 about 0.2 $M_\odot$ have been lost, the mass transfer rate decreases and the neutron
star enters a ``propeller phase'' during which the fast rotating neutron 
star will slow down. To avoid irradiation--driven mass loss from this system 
the external irradiation flux must be less than critical flux 
$\sim 1.6\times10^{10}$ erg$s^{-1}cm^{-2}$ (Podsiadlowski 1991). 
Let us estimate the value of the initial pulsar period $P_0$
to allow such a flux. Simple estimates show that $P_0$  
must be longer than 3.85$\times 10^{-3}$s, and the time elapsed from detachment 
of the Roche lobe is $\sim 3.2\times 10^9$ yrs. 
If $P_0$ $>$ 3.85$\times 10^{-3}$s then there is no irradiation driven mass
loss and the hydrogen burning is stationary, prolonging the cooling time. 
Since both  systems are in many parameters very similar to each 
other, only the white dwarf cooling time--scale is very different and there 
must be something which leads to the different cooling histories for the 
helium white dwarfs in these systems.  Different 
irradiation fluxes ($F_x$(J0751+1807) $\approx 10\times F_x$(J1012+5307)) 
may explain the discrepancy.

We would like to point to an important issue  neglected in many papers.
Usually the comparison with 
observed system  is made for one or several parameters (for 
example the cooling age of the white dwarf, $T_{eff}$ and so on) but not with 
all available data. For binary systems we must build a selfconsistent 
picture  with the right orbital parameters 
(which may be a very hard task, as in the case of the two systems 
discussed here), masses (if known) and cooling parameters of the white dwarf.

The effect of irradiation on the secular evolution of binaries containing
a millisecond pulsar has been discussed in a recent review paper by D'Antona
(1996). She has pointed out that one needs to distinguish the two main
effects of irradiation with following terms; i) illumination: structural
changes in the whole stellar structure due to irradiation, which may cause
the secondary to fill its Roche lobe again, ii) evaporation: direct driving
of mass loss from the secondary. Both cases give the same final result -- the
decrease of the mass of the hydrogen envelope.

We propose that for PSR J0751+1807 the irradiation by the pulsar causes
irradiation--driven mass loss from secondary, resulting in exhaustion of 
the hydrogen envelope, and as a consequence of this, a hydrogen shell 
burning stops. The irradiation--driven wind influences also the cooling 
phase, decreasing the cooling time--scale by a factor 1.7--4 (depending on 
how efficient this process is).

For the PSR J1012+5307 system the irradiative flux may be too small to cause
significant mass loss, and hydrogen shell burning will be maintained for a
long enough time to keep the white dwarf hot.

\section*{\sc Acknowledgements}
We would like to thank Prof. M. R\'o\.zyczka for help in improving the form
and text of the paper. We would like to thank referee Dr. F. D'Antona for 
constractive comments.  This work was supported in part by 
the Polish National Committee for Scientific Research under grant 
2--P03D--005--16 and 2--P03D--014--07, and by the Estonian SF grant 4338.

\end{document}